\documentclass[smallextended]{article}
\usepackage[latin2]{inputenc}

\usepackage{amsmath} 
\usepackage{amsthm, amssymb, latexsym} 
\usepackage{epsfig}
\usepackage[T1]{fontenc}
\usepackage{ae,aecompl}
\usepackage{graphicx}
\usepackage{float} 
\usepackage[all]{xy}
\usepackage{algorithmic}
\usepackage{algorithm}
\usepackage[noadjust]{cite}
\usepackage{subfigure}
\usepackage[percent]{overpic}

\usepackage{anysize}
\marginsize{3cm}{3cm}{3cm}{3cm} 
\floatname{algorithm}{Algoritmus}
\newtheorem{thm}{Theorem}

\newtheorem{all}{Statement}

\theoremstyle{definition}
\newtheorem{defi}{Definition}

\newcommand\blfootnote[1]{%
  \begingroup
  \renewcommand\thefootnote{}\footnote{#1}%
  \addtocounter{footnote}{-1}%
  \endgroup
}

\title{Indecomposable coverings with homothetic polygons}

\author{%
Istv\'an Kov\'acs}

\begin{document}
\maketitle

%

\blfootnote{This work has been supported by OTKA NN-102029.}

\begin{abstract}
We prove that for any convex polygon $S$ with at least four sides, or a concave one with no parallel sides, and any $m>0$, there is an $m$-fold
covering of the plane with homothetic copies of $S$ that cannot be 
decomposed into two coverings.
\end{abstract}

\section{Introduction}

Let ${\cal C}=\{\ C_i\ |\ i\in I\ \}$ be a collection of planar sets.
It is an {\em $m$-fold covering} if every point in the plane
is contained in at least $m$ members of $\cal C$. A $1$-fold covering is
simply called a {\em covering}.

\smallskip

A planar set $S$ is said to be cover-decomposable
if there is a constant $m=m(S)$ such that
every $m$-fold covering of the plane with translates of $S$
can be decomposed into two coverings.
J. Pach \cite{P80}, proposed the problem of determining all cover-decomposable
sets in 1980. He conjectured, that all planar convex sets are
cover-decomposable. The conjecture has been verified, in several steps, for
all convex polygons \cite{PT10} (see also \cite{P86}, \cite{TT07}).
However, very recently, P\'alv\"olgyi proved that the unit disc is not 
cover-decomposable \cite{P13}. His result holds also for convex sets with smooth
boundary.

The problem of determining cover-decomposable
sets has been generalized in many directions, see \cite{PPT13} for a
survey. 

A {\em homothetic} transformation is the composition of a translation and a scaling.
Keszegh and  P\'alv\"olgyi \cite{KP12} proved that any $12$-fold covering
of the plane with homothetic copies of a fixed triangle $T$ can be decomposed
into two coverings. 
In this note we prove that, with a few
possible exceptions, this result
cannot be extended to other polygons.

\begin{thm}\label{thm:main_thm}
Let $S$ be a convex polygon with at least four sides, or a concave
polygon with no parallel sides, and let $m>0$. 
There is an $m$-fold covering of the plane with homothetic copies of $S$ that cannot be 
decomposed into two coverings.
\end{thm}

For {\em convex} polygons we can keep the sizes of the  homothetic copies
``almost equal''.

\begin{thm}\label{egyforma_meret}
Let $S$ be a  convex polygon  with at least four sides, let $\varepsilon>0$
and $m>0$. 
There is a collection of  homothetic copies of $S$, each of them with scaling
factor
between $1-\varepsilon$ and $1+\varepsilon$, which forms an 
$m$-fold covering of the plane that cannot be 
decomposed into two coverings.
\end{thm}

Our method is based on the 
ideas of P\'alv\"olgyi \cite{P10}, \cite{P13}.

\section{Preparations}

Most of the papers about cover-decomposability investigate the problem in 
its {\em dual
form}. 

Suppose that ${\cal H}=\{\ S_i\ |\ i\in I\ \}$
is collection of {\em translates} of $S$
that form an $m$-fold covering of the plane.
For every $i\in I$, let $c_i$ be the center of gravity of $S_i$.
Let ${\cal H}'=\{\ c_i\ |\ i\in I\ \}$ be the set of the centers.
For any point $a$, let $-S(a)$ be a translate of $-S$ whose center of gravity
is $a$.
Then $a\in S_i$ if and only if $c_i\in -S(a)$.
Therefore, the collection ${\cal H}$
can be decomposed into two
coverings if and only if the points of the set
${\cal H}'$
can be colored with
two colors, such that  every translate of $S$ contains points of both
colors. This idea is originally due to J. Pach \cite{P86}.

If we have homothetic copies, then the dual version of the problem is not
equivalent to the original one. However, in this paper
we give a tricky definition
of the dual form.


Fix a coordinate system and let $o$ be the origin. 
If it does not lead to confusion,
for any point $p$, we denote its position vector 
$\overrightarrow{op}$ also by $p$. 
For any $\alpha$ real, set $S$, and point $p$, 
let 
$$\alpha\cdot S(p)=\{ \alpha\cdot x+p \ |\  x\in S\}.$$
The Minkowski sum of any convex polygons $S$ and $T$ 
is defined as $$S+T=\{ s+t \ |\  s\in S, t\in T\}.$$

\smallskip

Let $S$ be a fixed convex polygon of at least four sides, $o\in S$. 
It is well known \cite{S93} 
that for any $\alpha, \beta\ge 0$ 
$$\alpha\cdot S+\beta\cdot S=(\alpha+\beta )\cdot S.$$
As an easy consequence, we get the following statement.

\begin{all}\label{dualizalas}
Let $\alpha, \beta\ge 0$, $p, q\in\mathbb{R}^2$.
$(\alpha+\beta )\cdot S(p)$ contains $q$ if and only if 
$\alpha\cdot S(p)$ and 
$-\beta\cdot S(q)$ intersect each other.
\end{all}

First, for every pair $(k, l)$, we will construct a collection of homothetic
copies of $S$, 
${\cal X}_{k,l}$
and a collection of translates of $-S$, 
${\cal Y}_{k,l}$
with the property that 
for every red-blue coloring of the elements of ${\cal X}_{k,l}$, there is an
element of ${\cal Y}_{k,l}$ which intersects exactly $k$ elements, all of which are red (resp. exactly $l$ elements, all of which are blue).

Then we ``dualize'' this construction, for $m=k=l$,  as follows.
Replace each element of ${\cal X}_{m,m}$ by a larger homothetic copy, 
let ${\cal X}'_{m,m}$ be the new collection. 
Replace  each element of ${\cal Y}_{m,m}$ by a point, let
${\cal Y}'_{m,m}$ be the set of these points.

By Statement \ref{dualizalas},  ${\cal X}'_{m,m}$ and ${\cal Y}'_{m,m}$ have
the following property.

For every red-blue coloring of the elements of ${\cal X}'_{m,m}$, there is an
element (point) of ${\cal Y}'_{m,m}$ which is contained in exactly $m$ elements
of 
${\cal X}'_{m,m}$, all of which are of the same color.

So, for every $m$, ${\cal X}'_{m,m}$ forms a non-decomposable $m$-fold covering of
the points in ${\cal Y}'_{m,m}$.  Finally, we extend it to a non-decomposable
$m$-fold covering of the whole plane.

\section{Proof of Theorems \ref{thm:main_thm} and \ref{egyforma_meret}}

Let $S$ be a fixed convex polygon of at least four sides, $o\in S$. 
We say that $o$ is the {\em center} of $S$. We can assume that $S$ is
contained in the unit disc of center $o$. 
By definition, $-S$ denotes the reflection of $S$ about the origin.
Let  $v_1, v_2, \ldots , v_{n}$ be the vertices of $-S$, ordered clockwise. 
Indices are understood mod $n$, that is, $v_{n+1}$ means $v_1$.

\begin{defi}\label{wedgedef}
For every 
$i$, $1\le i\le n$, let $E^i$ denote the convex wedge whose apex is at the
origin and its bounding 
halflines are the translates of $\overrightarrow{v_iv_{i-1}}$ and  
$\overrightarrow{v_iv_{i+1}}$. $E^i$ is called the wedge that belongs to vertex $v_i$ of $-S$.
\end{defi}

\begin{figure}[H]
\begin{center}
  \begin{overpic}[height=55mm]{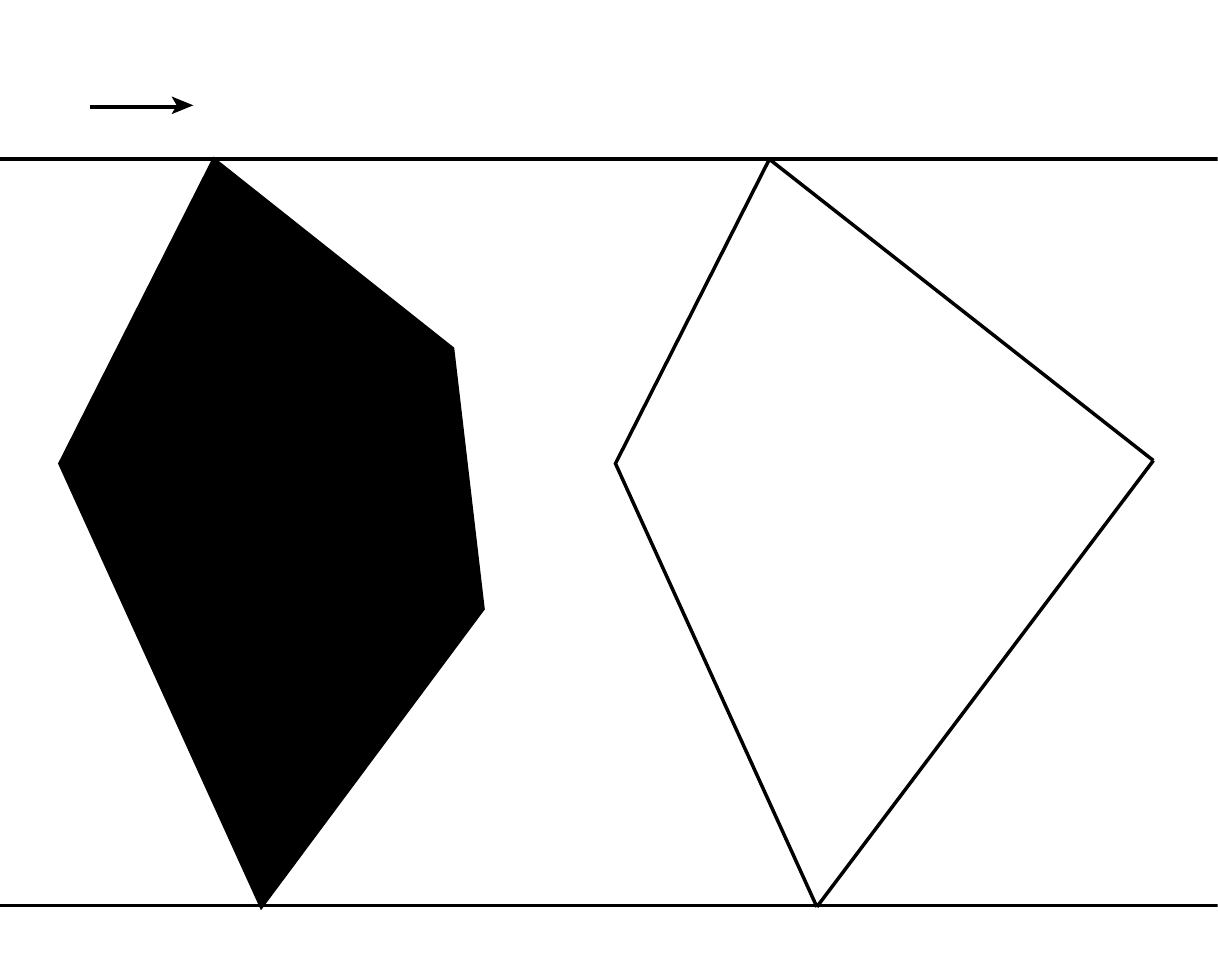} 
   \put(65,0){$v_b$}
   \put(62,68){$v_a$}
   \put(45,39){$v_l$}
   \put(96,38){$v_r$}
   \put(20,0){$v_b$}
   \put(18,68){$v_a$}
      \put(10,71){$d$}
   \put(52,56){$Q$}
      \put(7,56){$S$}
   \end{overpic}
\caption{$S$, $Q$ and the corresponding vertices.}\label{fig:minusS}
\end{center}
\end{figure}
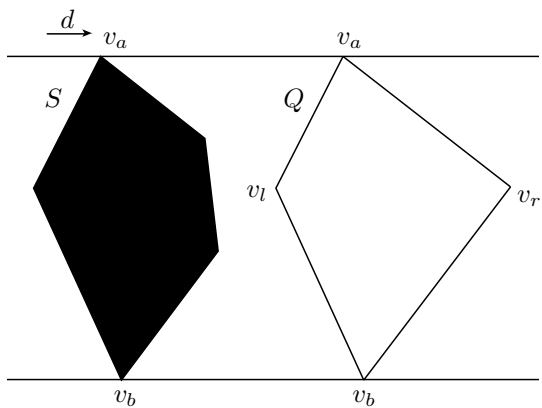

Choose a direction $d$ which is \textit{not parallel} to the sides of $S$, and the
two vertices, $v_a$  and $v_b$, where $S$ can be touched by a line parallel
to $d$, are not adjacent. Assume without loss of generality that $d$ is
horizontal, $v_a$ is the highest,
$v_b$ is the lowest vertex of $S$. Let $Q$ be a quadrilateral created from $S$ by extending the sides at $v_a$ and $v_b$. Let $v_r$ and $v_l$ be the rightmost and the leftmost vertices of $Q$, respectively. See Figure \ref{fig:minusS}. We can assume without loss of generality that $v_l$ is not lower than
$v_r$. Indeed, if $v_l$ is lower than
$v_r$, then we can apply a reflection of $S$ about the $y$-axis.
Let $\delta>0$ be a very small constant. 

For every pair $(k,l)$, $k, l\ge 1$, we will construct  a triple 
${\cal T}_{k,l}=({\cal X}_{k,l}, {\cal E}^a_{k,l},  {\cal E}^b_{k,l})$, where 
${\cal X}_{k,l}=\{\varepsilon_i\cdot S(p_i)\ |\ i\in I_{k,l}\}$,
$\varepsilon_i>0$, a collection of homothetic copies of $S$, 
${\cal E}^a_{k,l}=\{E^a(q_j)\ |\ j\in J^a_{k,l}\}$ and 
${\cal E}^b_{k,l}=\{E^b(r_j)\ |\ j\in J^b_{k,l}\}$ are collections of
translates of the wedges 
$E^a$ and $E^b$, respectively, for some  $I_{k,l},  J^a_{k,l}, J^b_{k,l}$ index sets.
${\cal T}_{k,l}$ will have the following properties. 

\smallskip

\noindent {\bf Property (1)} {\em For 
every red-blue coloring of the elements of ${\cal X}_{k,l}$, 
either there is an element of ${\cal E}^a_{k,l}$ which intersects exactly $k$ elements
of 
${\cal X}_{k,l}$, all of which are red, or there is an element of ${\cal E}^b_{k,l}$ which intersects exactly $l$ elements
of 
${\cal X}_{k,l}$, all of which are blue.}

\smallskip

\noindent {\bf Property (2)} {\em There is a disc $D_{k,l}$ of radius $\delta$
  which contains all apices of the wedges in 
${\cal E}^a_{k,l}$ and ${\cal E}^b_{k,l}$, and all elements of 
${\cal X}_{k,l}$.}

\smallskip

\begin{figure}[H]
\begin{center}
  \begin{overpic}[height=35mm]{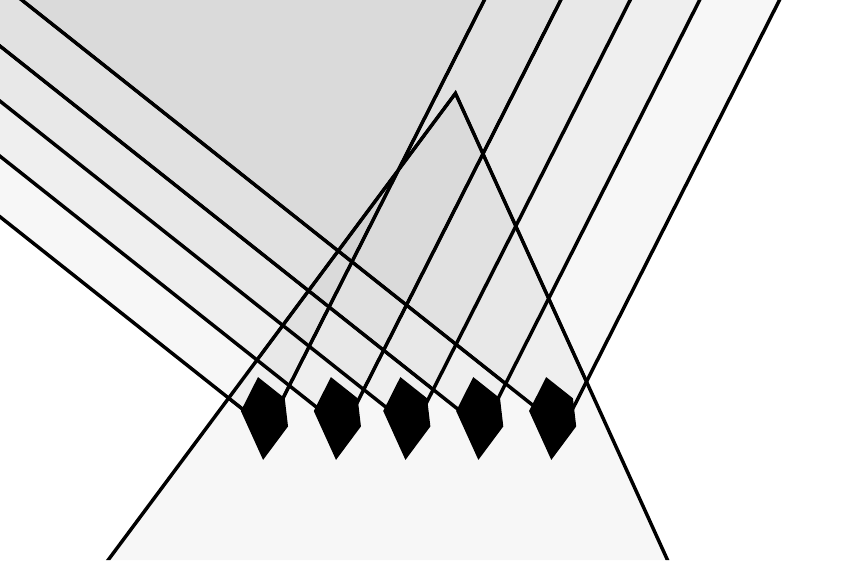} 
   \put(50,4){${\cal E}^a_{k,1}$}

      \put(30,56){${\cal E}^b_{k,1}$}
       \put(73,18){${\cal X}_{k,1}$}
   \end{overpic}
\caption{The construction of ${\cal T}_{k,1}$.}\label{fig:n1}
\end{center}
\end{figure}

First we define ${\cal T}_{k,1}$ and ${\cal T}_{1,l}$. For arbitrary $k$, let ${\cal X}_{k,1}$ be $k$ very small homothetic copies
of $S$, very close to each other on a horizontal line.
${\cal E}^a_{k,1}$ contains one translate of the wedge $E^a$ that intersects all $k$
homothetic copies, and ${\cal E}^b_{k,1}$ contains $k$  translates of the wedge $E^b$,
each intersects exactly one of the $k$
homothetic copies, but each intersects a different one. See Figure \ref{fig:n1}.
We define the triple ${\cal T}_{1,l}$ similarly, for any $l$. 

\begin{figure}[H]
\begin{center}
\includegraphics[height=60mm]{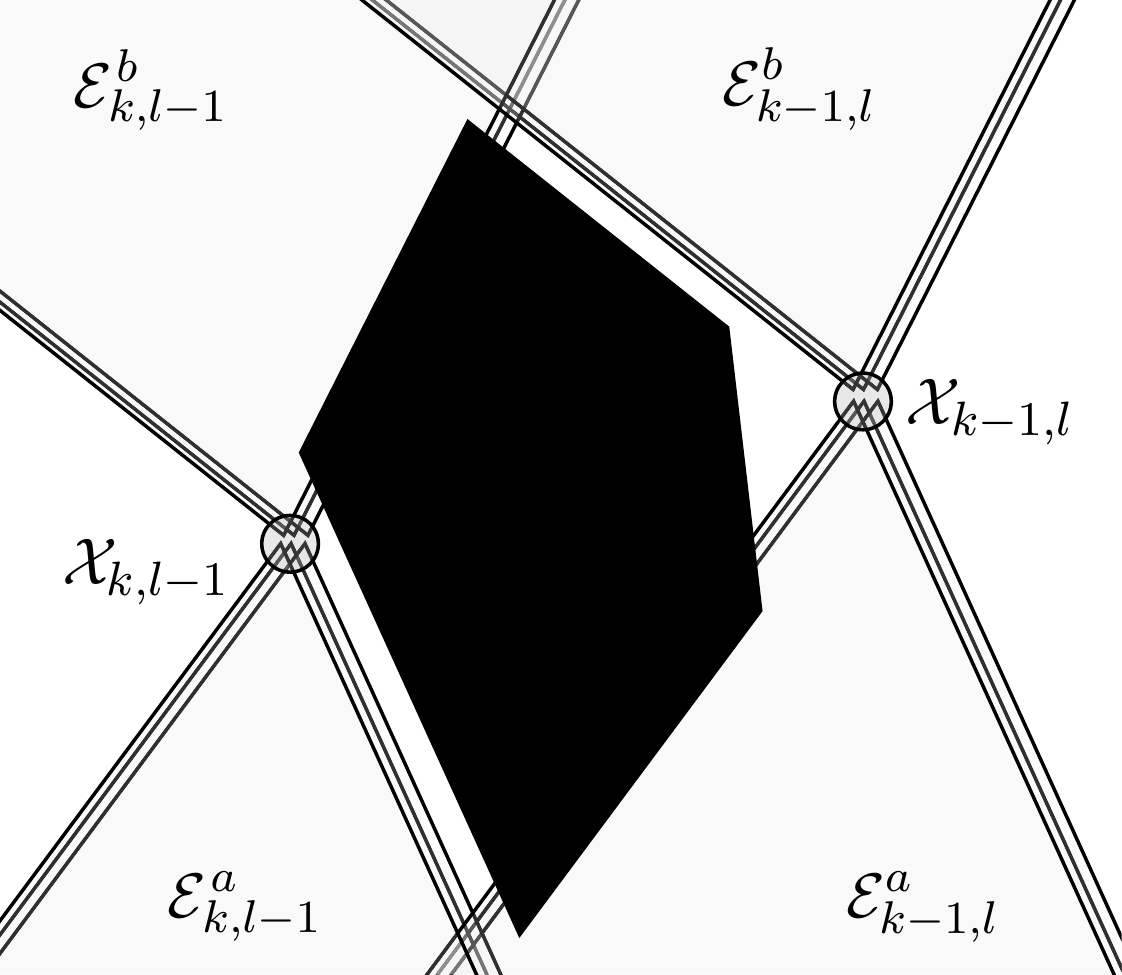} 
\caption{The induction step.}\label{fig:induction_step}
\end{center}
\end{figure}

Suppose now, that we have already defined 
${\cal T}_{k,l-1}$ and ${\cal T}_{k-1,l}$. 
Take a translate of ${\cal T}_{k,l-1}$ so that the center of $D_{k,l-1}$ is $(0,0)$, and a  
translate of ${\cal T}_{k-1,l}$ so that the center of  
$D_{k-1,l}$ is $(1, 3\delta )$. 
Place a suitable homothetic copy $S'=\varepsilon\cdot S$ of $S$ between 
points $(0,0)$ and $(1, 3\delta )$ such that
\begin{enumerate}
\item[(i)] $S'$ intersects all wedges in ${\cal E}^b_{k,l-1}$, and all wedges in 
${\cal E}^a_{k-1,l}$,
\item[(ii)] $S'$ does not  intersect any of the wedges in ${\cal E}^a_{k,l-1}$, and
any of the wedges in 
${\cal E}^b_{k-1,l}$.
\end{enumerate}
\noindent See Figure \ref{fig:induction_step}.
 
Let $${\cal X}_{k,l}={\cal X}_{k-1,l}\cup{\cal X}_{k,l-1}\cup\{S'\},$$
  $${\cal E}^a_{k,l}={\cal E}^a_{k-1,l}\cup{\cal E}^a_{k,l-1}, \ \ 
{\cal E}^b_{k,l}={\cal E}^b_{k-1,l}\cup{\cal E}^b_{k,l-1}.$$

Apply a suitable scaling, so that Property (2) is satisfied. 
We claim that Property (1) is also satisfied. 
Color the elements of 
${\cal X}_{k,l}$ by red and blue. Suppose that $S'$ is red. In the
subconfiguration that corresponds to  ${\cal T}_{k-1,l}$, either there is a
translate of $E^a$ 
that intersects exactly $k-1$ elements
of 
${\cal X}_{k-1,l}$, all of which are red, or there is a
translate of $E^b$ 
that intersects exactly $l$ elements
of 
${\cal X}_{k-1,l}$, all of which are blue.
In the first case, the corresponding translate of $E^a$ intersects exactly one 
more element of  ${\cal X}_{k,l}$, $S'$, and it is red, so we are done.
In the second case,  the corresponding translate of $E^b$
does not intersect any
other element of ${\cal X}_{k,l}$, so we are done again.
We can argue the same way, if $S'$ is colored blue.
Consequently, Property (1) is satisfied.

\smallskip

To obtain a non-decomposable $m$-fold covering, consider
${\cal T}_{m,m}=({\cal X}_{m,m}, {\cal E}^a_{m,m},  {\cal E}^b_{m,m})$.
${\cal X}_{m,m}=\{\varepsilon_i\cdot S(p_i)\ |\ i\in I_{m,m}\}$,
$\varepsilon_i>0$,  a collection of homothetic copies.

Replace each element of ${\cal E}^a_{m,m}$ (resp. ${\cal E}^b_{m,m}$) by a
translate of $-S$ such that its vertex $v_a$ (resp. $v_b$) moves to its apex.
We obtain a 
collection of translates of $-S$,  
${\cal Y}_{m,m}=\{-S(q_j)\ |\ j\in J_{m,m}\}$, 
with the property that 
for every red-blue coloring of the elements of ${\cal X}_{m,m}$, there is an
element of ${\cal Y}_{m,m}$ which intersects exactly $m$ elements
of 
${\cal X}_{m,m}$, all of the same color.

Let ${\cal X}'_{m,m}=\{(1+\varepsilon_i)\cdot S(p_i)\ |\ i\in I_{m,m}\}$,
a collection of homothetic copies of $S$, and let
${\cal Y}'_{m,m}=\{q_j\ |\ j\in J_{m,m}\}$, a collection of points. 
By Statement \ref{dualizalas},  
%
%
for every red-blue coloring of the elements of ${\cal X}'_{m,m}$, there is an
element (point) of ${\cal Y}'_{m,m}$ which is contained in exactly $m$ elements
of 
${\cal X}'_{m,m}$, all of the same color.

\smallskip

That is, ${\cal X}'_{m,m}$ forms a non-decomposable $m$-fold covering of the
points in
${\cal Y}'_{m,m}$. Moreover, for any $\varepsilon>0$, if we choose $\delta$
small enough, then the scaling factor of each member of ${\cal X}'_{m,m}$ is 
between $1-\varepsilon$ and $1+\varepsilon$. 



Now we extend  ${\cal X}'_{m,m}$ to a 
non-decomposable $m$-fold covering of the whole plane as follows.
We will add homothetic copies of $S$ to  ${\cal X}'_{m,m}$ that do not contain 
any point in ${\cal Y}'_{m,m}$, but each point in the plane will be covered at
least $m$ times. 
If we allow arbitrary small copies in the covering, then the extension is
trivial, since ${\cal Y}'_{m,m}$ is a finite point set. Just add {\em all}
homothetic copies of $S$ that do not contain any point of ${\cal Y}'_{m,m}$.

\begin{figure}
\begin{center}
\includegraphics[height=35mm]{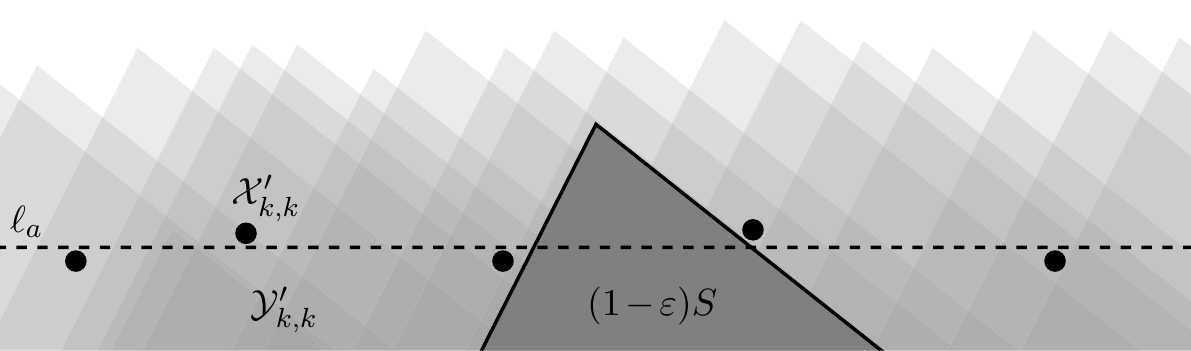} 
\caption{The points of type $a$  in ${\cal X}'_{m,m}$ are almost on the line $\ell_a$.}\label{fig:extend1}
\end{center}
\end{figure}

\begin{figure}[h!]
 \begin{center}
 \includegraphics[height=100mm]{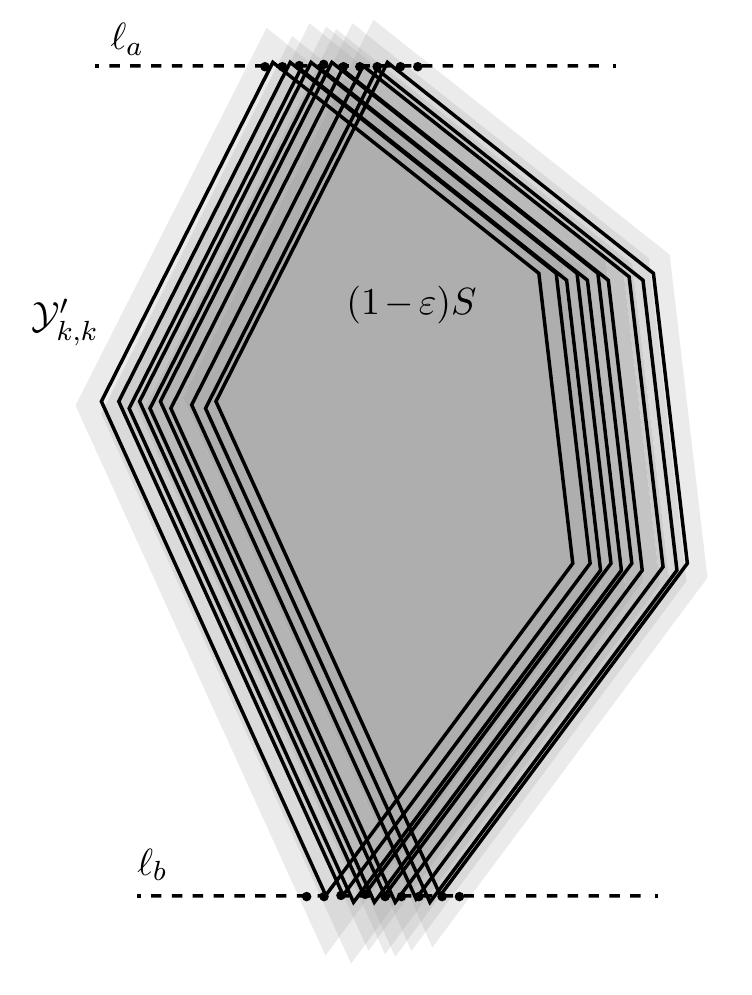} 
 \caption{Extending the cover by translates of $(1-\varepsilon)S$. }\label{fig:betomes}
 \end{center}
 \end{figure}

If we want to keep the sizes almost equal, we have to be more careful.
Points in  ${\cal Y}'_{m,m}$ are of two types, type $a$ (resp. type $b$) is the
set of those which come from a wedge in 
${\cal E}^a_{m,m}$ (resp. ${\cal E}^b_{m,m}$).
Observe, that any two points of the same type determine a line which is almost
horizontal.
In fact, we can take two horizontal lines, $\ell_a$ and $\ell_b$, at distance 
$Vert(v_av_b)$, such that all
points of type $a$ (resp. type $b$) are at distance at most $\delta$ from
$\ell_a$ (resp. $\ell_b$). See Figures \ref{fig:extend1} and \ref{fig:betomes}.

Add all translates of $(1-\varepsilon)S$ which avoid the points in 
${\cal  Y}'_{m,m}$. Now it is not hard to see that the resulting collection is
an $m$-fold covering of the whole plane, and by the 
construction of  ${\cal  X}'_{m,m}$ and ${\cal Y}'_{m,m}$, it is not
decomposable.
This concludes the proof of Theorem \ref{egyforma_meret}, 
and also the proof 
of Theorem \ref{thm:main_thm} in the special case when $S$ is convex. 

\smallskip

Now suppose that $S$ is concave with no parallel sides and let $m>0$. D. P\'alv\"olgyi \cite{P10} constructed a
collection 
${\cal X}'_{m,m}$ of {\em translates} of $S$, and a set ${\cal Y}'_{m,m}$ of points
such that 
${\cal X}'_{m,m}$
forms a non-decomposable $m$-fold covering of the
points in
${\cal Y}'_{m,m}$. 
Add {\em all}
homothetic copies of $S$ that does not contain any point of ${\cal Y}'_{m,m}$.
The resulting collection is clearly an $m$-fold covering of the 
plane, and just like in the previous argument, it is not
decomposable. This finishes the proof of Theorem~1.
\bigskip

\noindent {\bf Remark 1.}  The dual version of this problem is still open. 
Let $S$ be a polygon of at least
four sides. Is there an $m=m(S)$ with the following property?
Any point set 
${\cal P}$ can be colored with 
two colors such that if a 
homothetic copy of $S$ contains at least $m$ points of ${\cal P}$, 
then it
contains points of both colors. If $S$ is concave and has no parallel sides,
then the answer is NO to this question, even if we use only translates instead
of homothetic copies, by the result of P\'alv\"olgyi \cite{P10}.  

On the other hand, if $S$ is convex and 
we use only translates, then the answer is YES, by \cite{PT10}. 
If 
we do not allow arbitrarily 
large and arbitrarily small homothetic copies, then the answer is still YES, 
the proof in \cite{PT10} works also in this case.
But if we allow all homothetic copies, then the problem is unsolved.
See \cite{KP13} for related results.

\medskip

\noindent {\bf Remark 2.} We can define a hypergraph ${\cal H}_{k,l}$ to the
pair $({\cal X}_{k,l}, {\cal Y}_{k,l})$ in a natural way, 
elements of ${\cal X}_{k,l}$ correspond to the vertices and elements of 
${\cal Y}_{k,l}$ correspond to the hyperedges, 
a hyperedge contains a vertex if and only if the 
corresponding elements intersect each other. 
The same hypergraph was used by P\'alv\"olgyi in \cite{P10} and in \cite{P13}
to show that some concave polygons and the unit disc are not
cover-decomposable.
\medskip

\noindent {\bf Remark 3.} It was shown in \cite{PTT07}, that for every $m$,
there exists an  $m$-fold covering of the plane 
with axis-parallel rectangles that cannot be 
decomposed into two coverings.
We can slightly strengthen 
this result.

\begin{thm}\label{axis-parallel}
For any $m>0$, 
there is an $m$-fold covering of the plane with axis-parallel rectangles, each
with
unit horizontal side, 
that cannot be 
decomposed into two coverings.
\end{thm}

The proof is almost identical to the proof of Theorem \ref{thm:main_thm}. 
The main difference is that in the induction step, instead of a very small
copy of $S$, we add a very short vertical segment. We omit the details.

\medskip

\noindent {\bf Remark 4.} We believe that Theorem \ref{egyforma_meret}
can be extended to concave polygons with no parallel sides.

\section*{Acknowledgment}

I would like to thank my supervisor G\'eza T\'oth for all the help and  for
the many discussions. 
Without him this article would not have been completed.


\bibliography{ellenp_cikk_cite}
\bibliographystyle{abbrv}

\end{document}